\documentclass[onecolumn,aps,floatfix,pra,superscriptaddress,nofootinbib]{revtex4}

\usepackage{color}
\usepackage{graphicx}
\usepackage{bm}
\usepackage{amsmath}
\usepackage{hyperref}
\usepackage{enumerate}
\usepackage{upgreek}
\usepackage[subnum]{cases}
\usepackage{dsfont}

\newcommand{\tr}{ \text{tr} }

\newcommand{\pa}{ \partial }
\newcommand{\hb}{ \hbar }
\newcommand{\si}{ \sigma }
\newcommand{\om}{ \omega }

\newcommand{\ga}{ \gamma }
\newcommand{\la}{ \langle }
\newcommand{\ra}{ \rangle }
\newcommand{\del}{ \delta }
\newcommand{\al}{ \alpha }
\newcommand{\be}{ \beta }
\newcommand{\re}{ \text{Re} }
\newcommand{\im}{ \text{Im} }

\newcommand{\erf}{ \text{erf} }
\newcommand{\MB}{ \text{MB} }
\newcommand{\BE}{ \text{BE} }
\newcommand{\FD}{ \text{FD} }
\newcommand{\tot}{ \text{tot} }

\begin{document}

\title{Identical damped harmonic oscillators described by coherent states}
\author{S. V. Mousavi}
\email{vmousavi@qom.ac.ir}
\affiliation{Department of Physics, University of Qom, Ghadir Blvd., Qom 371614-6611, Iran}

\begin{abstract}

Some aspects of quantum damped harmonic oscillator (DHO) obeying a Markovian master equation are considered in the absence of thermal noise. The continuity equation is derived and Bohmian trajectories are constructed. As a solution of the master equation, we take a single coherent state and compute analytically the relative entropy of coherence, $C_r$, in the energy, position and momentum bases. Although $C_r$ is constant in both the position and the momentum bases, it is a decreasing function of time in the energy basis becoming zero at long times, revealing its role as the preferred basis. Then, quantum coherence is computed for a superposition of two coherent states, a cat state, and also a superposition of two cat states in the energy basis as a function of separation, in the complex plane, between the two superposed states. It is seen that the quantum coherence increases with this separation. Furthermore, quantum coherence of superposition is compared to that of decomposed states in the superposition. Finally, considering a system of two non-interacting DHOs, the effect of quantum statistics is studied on the coherence of reduced single-particle states, the joint detection probability and the mean square separation of particles. Our computations show that the single-particle coherence for antisymmetric states is always less than that of symmetric ones. Furthermore, boson anti-bunching and fermion bunching is seen in this open system. This behavior of bosons is the matter-wave analogue of photon anti-bunching seen in a modified Hanbury Brown-Twiss (HBT) interferometer.

\end{abstract}

\maketitle




\section{Introduction}

All physical systems are open in the sense that they cannot be completely isolated. The environment is there and affects the system of interest. 
This interaction results an entanglement between the physical system and its environment producing rapid
decay of interference terms, leading ultimately to the classical behavior at the macroscopic level. This is the story of decoherence \cite{Ho-bookchap-2009}. 
Different approaches have been used to study quantum open systems including time-dependent Hamiltonians, non-linear Schr\"{o}dinger equations, master equations for the reduced density matrix describing the system of interest, the functional integral approach and stochastic dynamics in Hilbert space \cite{BrPe-book-2002, Wi-book-2008, Caldeira-book-2014, NaMi-book-2017}.  
Practically, a simple effective description is given when coupling to the environment is taken into account through the inclusion of dissipative and stochastic terms in the dynamical equations describing the system \cite{BrPe-book-2002}. 

Physical systems subjected to analytical potentials can be approximated by a harmonic oscillator near the local minima of the potential energy function. So, this problem is of fundamental importance in physics. As a paradigm for dissipative systems, the quantum mechanics of the damped harmonic oscillator is generally used in formulating the quantum optics \cite{De-PA-1979, De-PR-1981}. Because of application in many areas of physics, the quantization of the damped harmonic oscillator has attracted a lot of attention over the years. See for instance \cite{BoDuTe-Pramana-1985, Kh-EPJP-2019, DeFu-PRA-2020, WeCa-PRE-2020, RJ-Pramana-2020}. Coherent states as eigenstates of the annihilation operator possesses the minimum uncertainty product and play a central role in quantum optics and quantum computation.  
By use of the Caldirola-Kanai Hamiltonian for the damped harmonic oscillator, exact coherent states have been constructed \cite{YeUmGe-PRA-1987}. The effect of dissipation on a superposition of two coherent states has been studied by a Markovian master equation, both in zero and non-zero temperatures, through damping of off-diagonal elements of the density matrix \cite{BrPe-book-2002, WaMi-PRA-1985}. It has been found decoherence occurs at a rate proportional to the separation between the two states. 
In this paper, we will have a novel look at the decoherence effect in DHO by considering reduction of the {\it quantum coherence} with dissipation.

%
%
Coherence as a quantum feature is related to the superposition principle revealing the wave-like nature of matter. Apart from fundamental importance, it is a resource for quantum technological applications widely used in quantum information processing \cite{BaCrPl-PRL-2014, StAdPl-RMP-2017}. 
%
Different measures have been proposed for quantum coherence of finite dimensional settings, such as the $l_1-$norm and relative entropy of coherence \cite{BaCrPl-PRL-2014} and skew information-based quantifier \cite{Gi-PRL-2014}. Quantification of coherence for infinite-dimensional systems has been done in \cite{ZhShYoFa-PRA-2016} where by requiring energy constraints, it has been found that the relative entropy of coherence works as a well-defined quantification of coherence in the infinite-dimensional systems but not the $l_1-$norm of coherence. Based on the relative entropy a measure of coherence has been provided for Gaussian states \cite{Xu-PRA-2016}. 
A general procedure has been proposed for quantifying the superposition amongst any complete set of quantum states, whether they are orthogonal or not \cite{TaVoKwJe-PRL-2017}. 
Even though coherence is rooted in the superposition principle, the coherence of a superposition state cannot be directly obtained from the coherence of the individual states being superposed. Coherence of superposed states has been discussed in \cite{Yuetal-ArXiv-2016, YuGaWeZh-SR-2017, LiLi-QIP-2016, LoBoFrSe-SR-2020}.
Interference visibility has been related to the quantum coherence \cite{Qu-PRA-2019}. Furthermore, relation between the effect of decoherence and the quantum coherence was the subject of some studies \cite{VeMiQu-PA-2019, MoHe-EPJP-2021}.

Quantum mechanically, identical particles are indistinguishable. Identical bosons, having an integer spin, obey the Bose-Einstein (BE) statistics while identical fermions, having a half-integer spin, obey the Fermi-Dirac (FD) statistics. Distinguishable particles obey Maxwell-Boltzmann statistics. The state of a composite system of identical particles must have a given symmetry under the exchange of particles, it must be (anti-)symmetric for identical (fermions) bosons. A theoretical study of the effect of quantum statistics on the joint detection probability of {\it free} identical particles for non-dissipative dynamics \cite{MaGr-EPJD-2014} and also dissipative one \cite{MoMi-EPJP-2020} has revealed that bunching and anti-bunching can occur for fermions and bosons, respectively. Furthermore, the role of quantum statistics has been investigated in the decay dynamics of a multi-particle state suddenly released from a confining potential \cite{MaGr-AP-2015}. Boson anti-bunching seen here is the matter-wave analogue of photon anti-bunching seen in a modified HBT interferometer \cite{HaZhZhGuQu-Opt-2021}.

In this work, our aim is to explore quantum correlations, through quantum coherence, in a system of DHO in one side and a system of two identical DHOs in the other side. Furthermore, due to different statistics in systems of identical particles, two important quantities are going to be evaluated, the mean square separation (MSS) between particles, and the simultaneous detection probability by an extended detector. For this goal, this manuscript is organized as follows. In section \ref{sec: dho}, dynamical equation and damped coherent states of the DHO is given. Section \ref{sec: coh} is devoted to the quantum coherence of a single coherent state, a superposition of two such states, a cat state, and a superposition of two cat states. Section \ref{sec: iden} analyses a system of two identical DHOs. Relative entropy of coherence for single-particle reduced states, MSS between particles and joint detection probability are also computed for different statistics. Section \ref{sec: conclude} includes the concluding remarks.

\section{Damped harmonic oscillator: dynamical equation and coherent states} \label{sec: dho}

Consider a simple harmonic oscillator (SHO) with frequency $\omega_0$. Neglecting vacuum energy, its non-dissipative evolution is generated by the system Hamiltonian $ H_S = \hb \om_0 a^{\dag} a $, where $ a^{\dag} $ and $ a $ are respectively creation and annihilation operators.
In the quantum optical limit and at zero temperature, the Schr\"{o}dinger picture master equation describing DHO reads \cite{BrPe-book-2002}
\begin{eqnarray}  
\frac{d}{d t} \rho(t) &=& \left( -i \om_0 - \frac{\ga_0}{2} \right) a^{\dag} a \rho(t) + \left( i \om_0 - \frac{\ga_0}{2} \right) \rho(t) a^{\dag} a  + \ga_0 a \rho(t) a^{\dag}
\label{eq: master*}
\end{eqnarray}
where $\ga_0$ denotes relaxation rate or damping constant. 
%
%
Note that $\rho(t)$ describing the state of this {\it open} system is obtained by taking partial trace over the heat bath, $\rho(t) = {\text{tr}}_B( \rho_{\tot}(t) )$, where the total density matrix $ \rho_{\tot}(t) $ evolves unitarily under time evolution operator $ U(t) $ and the sub-index $B$ stands for the heat {\it Bath}. Thus, one has \cite{BrPe-ArXiv-2003}
\begin{eqnarray}  \label{eq: rhoSt_red}
\rho(t) &=& {\text{tr}}_B( U(t)\rho_{\tot}(0)U^{\dag}(t)).
\end{eqnarray}
The master equation (\ref{eq: master*}) is the differential form of the equation (\ref{eq: rhoSt_red}) in the appropriate limit. Using the {\it scaled} unit  
\begin{numcases}~
x \rightarrow \sqrt{ \frac{\hb}{m \om_0} } x \\
p \rightarrow \sqrt{m\om_0 \hb} ~ p \\
t \rightarrow \frac{t}{ \om_0 } \\
\ga_0 \rightarrow \om_0 \ga_0. 
\end{numcases}
the master equation (\ref{eq: master*}) recasts
\begin{eqnarray}  
\frac{d}{d t} \rho(t) &=& \left( - i - \frac{\ga_0}{2} \right) a^{\dag} a \rho(t) + \left( i - \frac{\ga_0}{2} \right) \rho(t) a^{\dag} a  + \ga_0 a \rho(t) a^{\dag}
\label{eq: master}
\end{eqnarray}
where 
\begin{numcases}~
a = \frac{1}{ \sqrt{2} }( x + i p ) \label{eq: annihilation}
\\
a^{\dagger} = \frac{1}{ \sqrt{2} }( x - i p ) \label{eq: creation}
\end{numcases}
are respectively the annihilation and creation operators.

To solve (\ref{eq: master}), we take the initial state a superposition of two coherent states,
\begin{eqnarray}  \label{eq: psi0}
| \psi(0) \ra &=& \mathcal{N} ( c_{\alpha} | \alpha \ra + c_{\beta} | \beta \ra )
\end{eqnarray}
where, apart from a phase factor, the normalization constant $ \mathcal{N} $ is given by
\begin{eqnarray}  \label{eq: normal_const}
 \mathcal{N} &=& \frac{1}{\sqrt{ |c_{\al}|^2 + |c_{\be}|^2 + 2 \re \{c^*_{\al} c_{\be} \la \al | \beta \ra \} }}
\end{eqnarray}
The coherent state $ | \alpha \ra $, being eigenvector of the annihilation operator $a$ with the corresponding complex eigenvalue $\alpha$ is expressed versus eigenvectors of non-dissipative system Hamiltonian $H_S$ as 
\begin{eqnarray}  \label{eq: coh_state}
| \al \ra &=&  e^{-|\al|^2/2} \sum_{n=0}^{\infty} \frac{\al^n}{\sqrt{n!}} |n\ra .
\end{eqnarray}
Coherent states are not orthogonal
\begin{eqnarray}  \label{eq: coh_orth}
\la \al | \beta \ra &=& \exp \left[ - \frac{ |\al|^2 + |\be|^2 }{2} + \al^* \be \right]
\end{eqnarray}
however form a complete set 
\begin{eqnarray}  \label{eq: coh_complete}
\int \frac{d^2 \alpha}{ \pi } | \alpha \rangle \langle \alpha | &=& \mathds{1}.
\end{eqnarray}

Due to the linearity of (\ref{eq: master}), its solution with the initial condition (\ref{eq: psi0}) is given by \cite{BrPe-book-2002}
\begin{eqnarray}  
\rho(t) &=& \mathcal{N}^2 \bigg\{ |c_{\al}|^2 | \al(t) \ra \la \al(t) | + |c_{\be}|^2 | \be(t) \ra \la \be(t) | + c_{\al} c^*_{\be} f(t) | \al(t) \ra \la \be(t) | 
\nonumber \\
&\qquad & \qquad + c_{\be} c^*_{\al} f^*(t) | \be(t) \ra \la \al(t) | \bigg\}
\label{eq: rhoSt}
\end{eqnarray}
where,
\begin{numcases}~
\al(t) = \al e^{-(i + \ga_0/2)t} \label{eq: alt}
\\
\be(t) = \be e^{-(i + \ga_0/2)t} \label{eq: bet}
\\
f(t) = \la \be | \al \ra^{1-e^{-\ga_0 t}} \label{eq: ft}
\end{numcases}
Natural logarithm of the modulus of the function $f(t)$ has been identified as the {\it decoherence function} \cite{BrPe-book-2002} which from its short time behaviour, $\ga_0 t \ll 1$, a decoherence time $\uptau_D$ has been defined through $ \uptau_D = \uptau_R \frac{2}{|\al - \be|^2} $ where $ \uptau_R = \ga_0^{-1} $ is the relaxation time and $ |\al - \be| $ denotes the distance of the initial states in the complex plane.

By using Eqs. (\ref{eq: alt}) and (\ref{eq: bet}) in Eq. (\ref{eq: coh_orth}) one obtains 
%
\begin{eqnarray}  \label{eq: coh_orth_t}
\la \al(t) | \be(t) \ra &=& \la \al| \be \ra^{e^{-\ga_0 t}} .
\end{eqnarray}
%


Using Eqs. (\ref{eq: annihilation}) and (\ref{eq: creation}) the master equation (\ref{eq: master}) can be written in the position representation as
\begin{eqnarray} 
\frac{\pa }{\pa t} \rho(x, x', t) &=& \frac{1}{i} \left[ -\frac{1}{2} \left( \frac{\pa^2}{ \pa x^2 } - \frac{\pa^2}{ \pa x'^2 } \right) + \frac{1}{2} (x^2-x'^2) \right] \rho(x, x', t) 
\nonumber \\
& & + \frac{\ga_0}{2} \left[ 1 + \left( x \frac{\pa}{\pa x'} + x' \frac{\pa}{\pa x} \right) - \frac{1}{2}(x-x')^2
+ \frac{1}{2} \left( \frac{\pa}{\pa x} + \frac{\pa}{\pa x'} \right)^2
\right] \rho(x, x', t) 
\nonumber \\
\label{eq: master_x}
\end{eqnarray}
where $ \rho(x, x', t) = \la x | \rho(t) | x' \ra $. 
In the new coordinates
\begin{numcases}~
R = \frac{x+x'}{2} \label{eq: cm} \\
r = x - x'  \label{eq: rel} 
\end{numcases}
Eq.(\ref{eq: master_x}) can be written as
\begin{eqnarray} \label{eq: mastereq_Rr}
\frac{\pa \rho}{\pa t} + \frac{\pa j}{\pa R} + \left[ - i~ r R + \frac{\ga_0}{2} 
\left( r \frac{ \pa }{\pa r} + \frac{1}{2} r^2 \right) \right] \rho &=& 0
\end{eqnarray}
where we have introduced the probability current density matrix
\begin{eqnarray} \label{eq: cur_den_mat}
j(R, r, t) &=& - \left[ i \frac{\pa}{\pa r} + \frac{\ga_0}{2} \left( R + \frac{1}{2} \frac{\pa}{\pa R} \right) \right] \rho(R, r, t) .
\end{eqnarray}
From this equation, one can construct the continuity equation 
\begin{eqnarray} \label{eq: con_non}
\frac{\pa}{\pa t} P(x, t) + \frac{\pa}{\pa x} J(x, t)  &=& 0 ,
\end{eqnarray}
where
\begin{numcases}~
P(x, t) = \int dx' ~ \del(x-x') ~ \rho(x, x', t) = \rho(R=x, r=0, t)  \\
J(x, t) = \int dx' ~ \del(x-x') ~ j(x, x', t) = j(R=x, r=0, t)
\end{numcases}
are respectively the probability density (PD) and the probability current density (PCD). Then, one can construct Bohmian trajectories $ x(x^{(0)}, t) $, $x^{(0)}$ being the initial position of the Bohmian particle, via integrating the guidance equation \cite{Holland-book-1993}; 
\begin{eqnarray} \label{eq: guidance}
\frac{dx}{dt} &=& \frac{J(x, t)}{P(x, t)} \bigg|_{x = x(x^{(0)}, t)} .
\end{eqnarray}

Coherent state wavefunction in the coordinate space reads
\begin{eqnarray}  
\psi_{\al}(x, t) &=& \psi_{\al(t)}(x, 0) 
= \frac{1}{ \pi^{1/4} }
\exp \left[ - \frac{ 1 }{ 2 } \left( x - \sqrt{2} \al_r(t) \right)^2
+ i ~ \sqrt{2} \al_i(t) ~ x ~+ ~i  \al_r(t) \al_i(t)
 \right] 
 \nonumber \\
 \label{eq: coh_wf}
\end{eqnarray}
where $ \al(t) $ is given by (\ref{eq: alt}); and $ \al_r(t) $ and $ \al_i(t) $ are respectively the real and imaginary parts of $ \al(t) $ i.e., $ \al_r(t) = \re\{\al(t)\} $ and $ \al_i(t) = \im\{\al(t)\} $.
%
One can easily check that $ \rho(x, x', t) = f(t) \psi_{\al}(x, t) \psi^*_{\be}(x', t) $ fulfils equation (\ref{eq: master_x}).

Momentum-space wavefunction of the coherent state is obtained by the Fourier transform of (\ref{eq: coh_wf}) which reads 
\begin{eqnarray}  
\phi_{\al}(p, t) &=&  \phi_{\al(t)}(p, 0) =
\frac{1}{ \pi^{1/4} }
\exp \left[ -\frac{ [ p - i \sqrt{2} ~ \al(t) ]^2 }{2} - \al_r(t) ( \al_r(t) - i \al_i(t) )
 \right] 
 \nonumber \\
  \label{eq: coh_wf_mom}
\end{eqnarray}

In the following, evolution of the state (\ref{eq: psi0}) with $c_{\al} =1, c_{\be}=0$ will be denoted by $\rho_1(t)$ and that of $c_{\al} =1= c_{\be}$ by $\rho_2(t)$ i.e.,
\begin{numcases}~
\rho_1(0)  = | \al \ra \la \al | \label{eq: rho01}
\\
\rho_2(0)  = \mathcal{N}^2 (| \al \ra +  | \be \ra)( \la \al | +  \la \be| ) \label{eq: rho02}
\end{numcases}
Noting (\ref{eq: rhoSt}), one sees that the state $\rho_1$ remains pure over the time.

\section{ Quantum coherence } \label{sec: coh}

Different measures of quantum coherence have been introduced \cite{BaCrPl-PRL-2014}; among them are $l_1$-norm and relative entropy. It has been proved that for infinite dimensional vector spaces, only relative entropy provides an acceptable measure of quantum coherence \cite{ZhShYoFa-PRA-2016}. 
Coherence of superposition and its relation to that of decomposed states in the superposition is also an important issue \cite{Yuetal-ArXiv-2016, YuGaWeZh-SR-2017, LiLi-QIP-2016, LoBoFrSe-SR-2020}.
%

\subsection{ Coherence of a coherent state with dissipation }

Consider the initial state of the system to be the coherent state (\ref{eq: rho01}). This state evolves under (\ref{eq: master}) to the state $ \rho_1(t) = |\al(t) \ra \la \al(t) | $ with $ \al(t) $ given by (\ref{eq: alt}). Given a particular basis $ \{ | i \ra \} $, the relative entropy of coherence of a state $ \rho $ is given by
\begin{eqnarray} \label{eq: rel.ent}
C_r(\rho) &=& S(\rho_d) - S(\rho)
\end{eqnarray}
where $ S(\rho) = - \tr( \rho \ln \rho ) $ is the von Neumann entropy of the state and $ \rho_d $ denotes the state obtained from $ \rho $ by deleting all off-diagonal elements in
the basis of question. For our pure state $ \rho_1(t) $ where the von Neumann entropy is zero, the definition (\ref{eq: rel.ent}) in the basis $ \{ | n \ra_{n=0}^{\infty} \} $ i.e., eigenvectors of $H_S$, is written as 
\begin{eqnarray} 
C_r(\rho_1(t)) &=& S( \rho_{1,d}(t) ) = - \sum_{n=0}^{\infty} \rho_{1,nn}(t) \ln( \rho_{1,nn}(t) )
\end{eqnarray}
where, $ \rho_{1,nn}(t) = |\la n | \al(t) \ra|^2 $. Thus, from $ \la n | \al(t) \ra = \dfrac{[\al(t)]^n}{ \sqrt{n!} } e^{-|\al(t)|^2/2} $ and (\ref{eq: alt}) one obtains
\begin{eqnarray}
C_r(\rho_1(t)) &=& |\al(t)|^2 ( 1 - \ln |\al(t)|^2  ) + e^{-|\al(t)|^2} \sum_{n=0}^{\infty} |\al(t)|^{2n} 
\frac{ \ln n! }{ n! }
\nonumber \\
&=&
e^{-\ga_0 t} |\al|^2 ( 1 - \ln|\al|^2 - \ga_0 t ) + e^{ e^{-\ga_0 t} |\al|^2 }
\sum_{n=0}^{\infty} (e^{-\ga_0 t} |\al|^2)^n \frac{ \ln n! }{ n! }.  
\nonumber \\
\label{eq: Cr_al}
\end{eqnarray}
This equation shows that for the non-dissipative dynamics, $ \ga_0 = 0 $, quantum coherence is independent of time. With dissipation, quantum coherence becomes zero at long times, $\ga_0 t \gg 1$. This means that in this limit the state is decohered. 
Note that for the case $ \al = 0 $ the state $ \rho_1(t) $ becomes $ |0\ra \la  0 |$,  $|0\ra$ being the ground state of SHO. Thus one has $ \rho_{1, nn} = \del_{n0}$ and $ C_r = - \sum_n \del_{n0} \ln \del_{n0} = 0 $ which is an expectable result noting that the state $ |0\ra \la  0 | $ is not a superposition state in the energy eigenbasis. According to the relation $ |\al \ra = \sum_{n} |n\ra \dfrac{\al^n e^{-|\al|^2/2} }{ \sqrt{n!} }  $ as $|\al|$ increases more terms of the summation contribute and thus one expects coherence, a measure of superposition, increases with this quantity.

Using Eqs. (\ref{eq: coh_wf}) and (\ref{eq: coh_wf_mom}) one obtains 
\begin{eqnarray}
C_r(\rho_1(t))\bigg|_{\text{position rep.}} &=& \frac{1 + \ln(\pi)}{2} 
\\
C_r(\rho_1(t))\bigg|_{\text{momentum rep.}} &=& \frac{1 + \ln(\pi)}{2} 
\end{eqnarray}
which show that coherency of the state in both position and momentum representations are time-independent. This means, this state will not be decohered in these representations. This point can also be seen by analysing the matrix elements of the density operator $\rho_1$; non-diagonal elements don't become zero at long times.  
From the above analysis one finds that the preferred basis is the energy basis \cite{Ve-PRA-1994}.

\subsection{ Coherence of superpositions of non-dissipative coherent states  }

\begin{figure}  
\centering
\includegraphics[width=12cm,angle=-0]{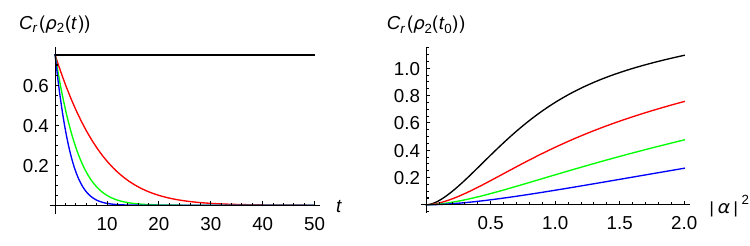}
\caption{(Color online)
Relative entropy of quantum coherence $ C_r(\rho_2(t)) $ for the superposition of two coherent states with $ \al = - \be = 1 $ versus time (left panel) and versus $ |\al|^2 $ for the given time $ t_0 = 5 $ (right panel) for different values of damping constant $ \ga_0 = 0 $ (black), $ \ga_0 = 0.1 $ (red), $ \ga_0 = 0.2 $ (green) and $  \ga_0 = 0.3 $ (blue).
}
\label{fig: crelcoh} 
\end{figure}

Even though coherence is rooted in the superposition principle, the coherence of a superposition state cannot be directly obtained from the coherence of the individual states being superposed. 
In this section we consider quantum coherence of the superposition state (\ref{eq: rho02}). 
Without dissipation, this state remains a pure one, $ | \rho_2(t) \ra = \mathcal{N}^2 ( | \al(t) \ra + | \be(t) \ra )( \la \al(t) | +  \la \be(t) | ) $ where $\al(t)$ and $\be(t)$ are given by eqs. (\ref{eq: alt}) and (\ref{eq: bet}) with $ \ga_0 =0 $. For $ \be = - \al $, that is superposition of two same-amplitude coherent states, one obtains
\begin{eqnarray}
C_r( \rho_2(t) ) &=& \ln( 2 \cosh|\al|^2 ) - |\al|^2 (\ln|\al|^2) (\tanh |\al|^2)
\nonumber \\
&-& \frac{1}{2 \cosh|\al|^2} \sum_{n=0}^{\infty} |\al|^{2n} \frac{ 1 + (-1)^n }{n!}  \ln\frac{ 1 + (-1)^n }{n!} 
 \label{eq: Cr_albe}
\end{eqnarray}
which is independent of time. Note that according to (\ref{eq: coh_wf}), $ |\al|^2 $ measures the square separation between two superposed wavepackets. As before, for $ \al = 0 $ this state is just the ground state and the quantum coherence is zero as it must be because we are working in energy eigen-basis. 
In the left panel of Fig. \ref{fig: crelcoh}, we have plotted the quantum coherence of this cat state versus time for different values of the relaxation rate. As one expects, the quantum coherence is constant for the non-dissipative dynamics while decreases with time for a given damping constant. Furthermore, dissipation diminishes the quantum coherence in a given time. In the right panel of this figure, relative entropy of quantum coherence has been plotted versus $ |\al|^2 $ at the given time $t_0=5$. 

%
%
\begin{figure}  
\centering
\includegraphics[width=12cm,angle=-0]{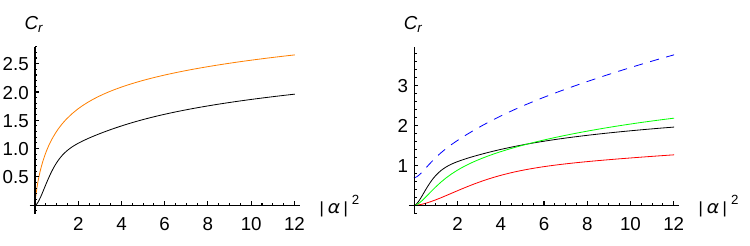}
\caption{(Color online)
Variation of relative entropy of quantum coherence with $ |\al|^2 $ for non-dissipative dynamics for a single coherent state, $ | \al \ra $ (orange curve), the cat state $ | \Phi \ra = N_{\Phi} ( | \al \ra + | - \al \ra ) $ (black curves), the cat state $ | \Psi \ra = N_{\Psi} ( | \al/2 \ra + | - \al/2 \ra ) $ (red curve) and the superposition of these two cat states $ | T \ra = N_T \frac{1}{\sqrt{2}} ( | \Phi \ra + | \Psi \ra ) $ (green curve). Dashed blue curve depicts the right side of the non-equality (\ref{eq: inequlity2}).
}
\label{fig: one&twocatNondis} 
\end{figure}

%
%
Coherence of superposed states has been discussed in \cite{Yuetal-ArXiv-2016, YuGaWeZh-SR-2017}. In Ref. \cite{Yuetal-ArXiv-2016} an inequality relation has been derived which contains coherence of both superposition state and decomposed in superposition states. 
According to eq. (15) of this reference  we should have
\begin{eqnarray} \label{eq: noneq1}
\frac{1}{2 \mathcal{N}^2} C_r( \rho_2(t) ) & \leq & 2 \left( \frac{1}{2} C_r( \rho_1(t) ) + \frac{1}{2} C_r( \rho_1(t) |_{\al \rightarrow - \al} ) + \ln(2) \right).
\end{eqnarray}
From eq. (\ref{eq: Cr_al}) one has $ C_r( \rho_1(t) |_{\al \rightarrow - \al} ) = C_r( \rho_1(t) ) $. Then, using $ \mathcal{N} = 1 / \sqrt{ 2( 1 + e^{-2\al^2} ) } $ one obtains an upper bound for the quantum coherence of the superposed state $ \rho_2(t) $
\begin{eqnarray} \label{eq: upbound}
 C_r( \rho_2(t) ) & \leq & \frac{1}{ e^{-|\al|^2} \cosh|\al|^2 } [ C_r( \rho_1(t) + \ln(2)]  .
\end{eqnarray}
Numerical computations (left panel of figure \ref{fig: one&twocatNondis}) show that $  C_r( \rho_2(t) ) \leq  C_r( \rho_1(t) ) $, with the equality holding for $ \al = 0 $. 
Since $ e^{-|\al|^2} \cosh|\al|^2 \leq 1 $, the upper bound in (\ref{eq: upbound}) is larger than $ C_r( \rho_1(t) ) $. This means that inequality $  C_r( \rho_2(t) ) \leq  C_r( \rho_1(t) ) $ is tighter than the inequality (\ref{eq: noneq1}), at least in our case.

We now consider a superposition of two cat states;
\begin{eqnarray}
| T \ra &=& \mathcal{N}_{T} \left( \frac{1}{ \sqrt{2} } | \Phi \ra + \frac{1}{ \sqrt{2} } | \Psi \ra   \right)
\end{eqnarray}
where
\begin{numcases}~
| \Phi \ra = \mathcal{N}_{\Phi} \left( | \al \ra + | - \al \ra   \right) \\
| \Psi \ra = \mathcal{N}_{\Psi} \left( \bigg | \frac{\al}{2} \bigg \ra + \bigg | - \frac{\al}{2} \bigg \ra  \right) 
\end{numcases}
to see the relation between coherence of these states i.e., $ C_r( | T \ra \la T | ) $, $ C_r( | \Phi \ra \la \Phi | ) $ and $ C_r( | \Psi \ra \la \Psi | ) $.
Again from Eq. (15) of \cite{Yuetal-ArXiv-2016} one obtains the inequality
\begin{eqnarray} \label{eq: inequlity2}
C_r( | T \ra \la T | ) & \leq & \mathcal{N}_{T}^2 [ C_r( | \Phi \ra \la \Phi | + C_r( | \Psi \ra \la \Psi | + 2 \ln(2) ].
\end{eqnarray}

Variation of quantum coherence with $ |\al|^2 $ has been depicted in the right panel of figure \ref{fig: one&twocatNondis} for cat states $ | \Phi \ra $ and $ | \Psi \ra $; and the superposition $ | T \ra $ of these two cat states. Dashed blue curve displays the right side of the non-equality (\ref{eq: inequlity2}).
%
%
%
%

%
%
%
\begin{figure}
\centering
\includegraphics[width=12cm,angle=-0]{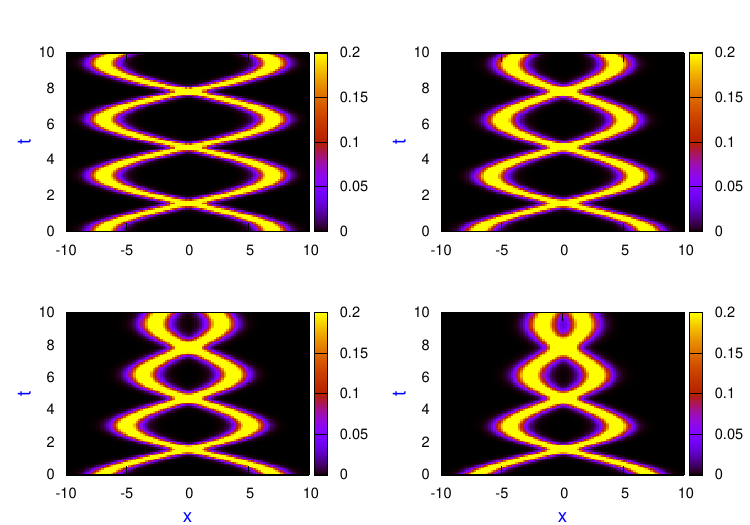}
\caption{
Density plots of the probability density (\ref{eq: rhoSt}) for the damping constant $\ga_0=0$
(left top panel), $ \ga_0 = 0.1 $ (right top panel) $\ga_0=0.2$ (left bottom panel), $ \ga_0 = 0.3 $ (right bottom panel).  
Parameters are chosen to be $\al = - \be = 7/\sqrt{2} $ and $c_{\al}=c_{\be}= 1$. 
} 
\label{fig: pd_sup} 
\end{figure}
\begin{figure}
\centering
\includegraphics[width=12cm,angle=-0]{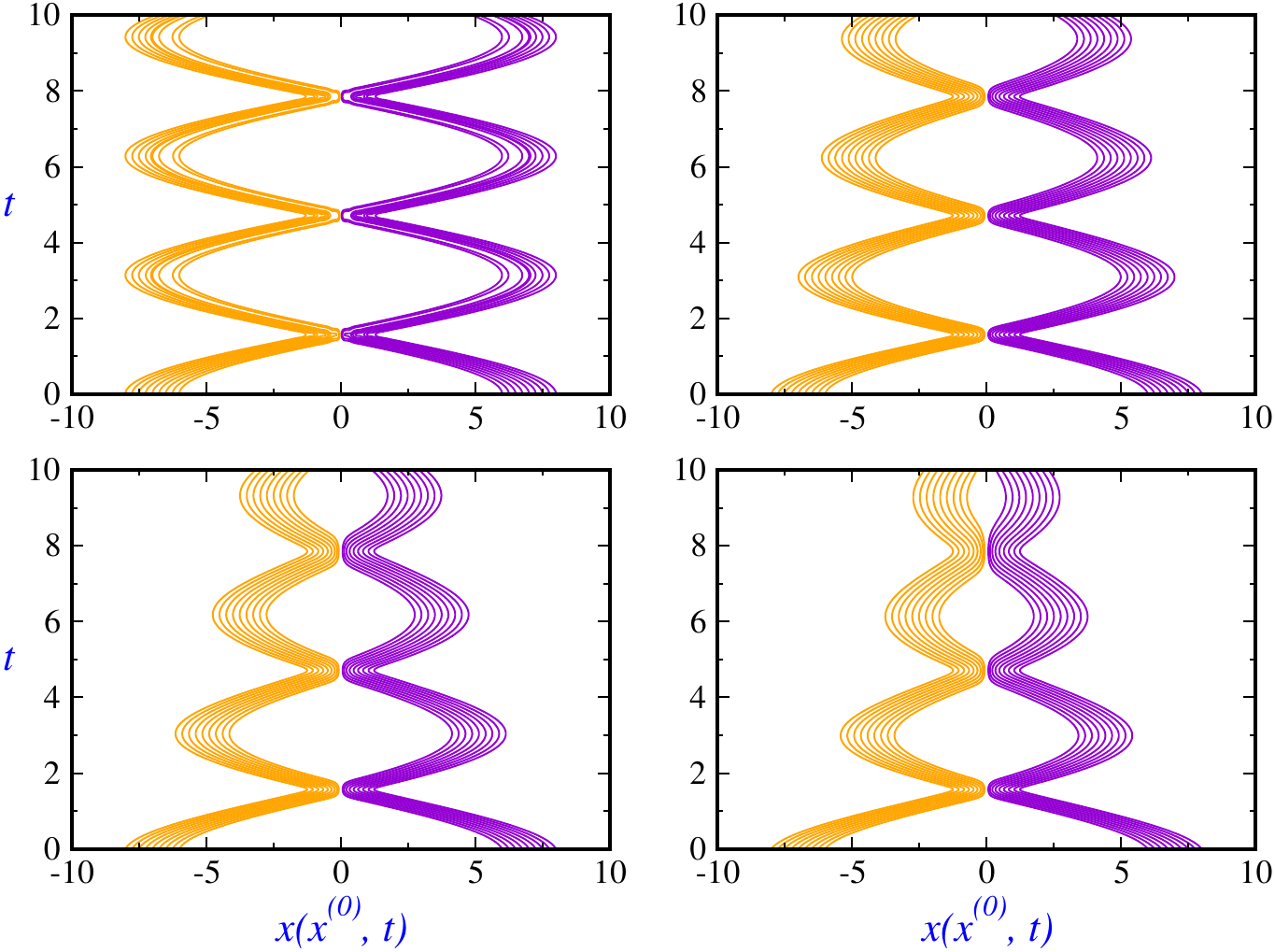}
\caption{
A selection of Bohmian trajectories corresponding to the figure \ref{fig: pd_sup}. 
} 
\label{fig: trajs} 
\end{figure}

%
%
In figure \ref{fig: pd_sup} we have plotted density plots of the position-space probability density for the superposed state (\ref{eq: rhoSt}) for $\al = - \be = 7/\sqrt{2} $ and $c_{\al}=c_{\be}= 1$ for different values of relaxation rate. This figure shows that the interference pattern is gradually washed out with increasing damping rate. Corresponding Bohmian trajectories has been plotted in figure \ref{fig: trajs}. Bohmian trajectories reflect when the superposed wave packets collides at the origin, remaining in the wavepacket where they started at. 

%
%
%
%
%

\section{ Identical quantum damped harmonic oscillators } \label{sec: iden}

\begin{figure} 
\centering
\includegraphics[width=12cm,angle=-0]{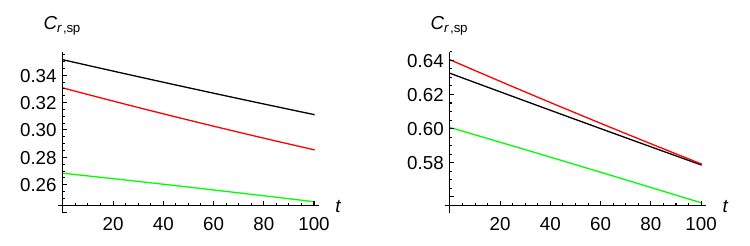}
\caption{(Color online)
Relative entropy of coherence for the reduced single-particle states (\ref{rhopm-sp}) and \ref{rhoMB-sp}) versus time for $ \ga_0 = 0.001 $ for $ \al = -\be = \frac{1}{\sqrt{2}} $ (left panel) and $ \al = -\be = 1 $ for different statistics: MB (black curves), BE (red curves) and FD (green curves).
}
\label{fig: crrhosptime} 
\end{figure}

\begin{figure} 
\centering
\includegraphics[width=12cm,angle=-0]{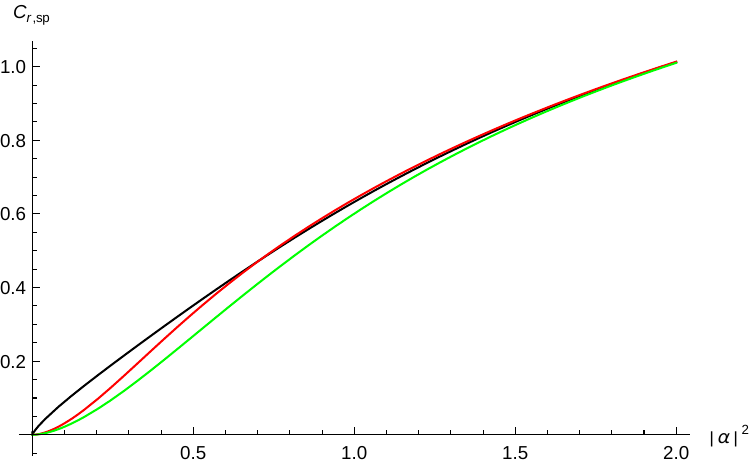}
\caption{(Color online)
Relative entropy of coherence for the reduced single-particle states (\ref{rhopm-sp}) and \ref{rhoMB-sp}) versus $ |\al|^2 $ for the non-dissipative dynamics, $ \ga_0 = 0 $, for different statistics: MB (black curves), BE (red curves) and FD (green curves).
}
\label{fig: crrhospalpha2} 
\end{figure}

Consider a system of two {\it independent non-interacting} harmonic oscillators which are described by density operators $ \rho_1 $ and $ \rho_2 $, respectively. Then the combined system will be described by the product state $ \rho_1 \otimes \rho_2 $. Writing the one-particle master equation (\ref{eq: master*}) as $ \dot{\rho}(t) = \mathcal{L} \rho(t) $, 
$ \mathcal{L} $ being the linear operator appearing in this equation, one can easily see that the time evolution for the combined system is given by
\begin{eqnarray} \label{eq: master_2p}
\frac{\pa}{\pa t} ( \rho_1 \otimes \rho_2 ) &=&  
( \frac{\pa}{\pa t} \rho_1 ) \otimes \rho_2 + \rho_1 \otimes \frac{\pa}{\pa t} \rho_2 =
( \mathcal{L}_1 + \mathcal{L}_2 ) ( \rho_1 \otimes \rho_2 )
\end{eqnarray}
Now, suppose these DHOs are identical and spinless. 
For such a system, the state of the system must have a given symmetry, it must be (anti)-symmetric under the exchange of particles for identical (fermions) bosons. 
We take the initial state to be the pure state 
\begin{eqnarray} \label{eq: psipm0}
| \Psi_{\pm}(0) \ra &=& \mathcal{N}_{\pm} ( | \al \ra_1 \otimes | \be \ra_2 \pm | \be \ra_1 \otimes | \al \ra_2 ) , \quad  \mathcal{N}_{\pm} = \frac{1}{ \sqrt{2(1\pm|\la \al | \be \ra |^2)} }
\end{eqnarray}
%
%
%
where the plus (minus) sign refers respectively to bosons (fermions). It should be emphasized that we have neglected the spin degrees of freedom and considered only the coordinate space state. Otherwise, one can think of spin-less particles. 
Under the {\it linear} evolution equation (\ref{eq: master}), time dependent density matrix is obtained as follows
\begin{eqnarray} 
\rho_{\pm}(t) 
&=& \mathcal{N}^2_{\pm} \bigg( | \al(t) \ra_1 {~}_1\la \al(t) | \otimes | \be(t) \ra_2 {~}_2\la \be(t) | 
+ | \be(t) \ra_1 {~}_1\la \be(t) | \otimes | \al(t) \ra_2 {~}_2\la \al(t) |
\nonumber\\
& \qquad & \qquad \pm |f(t)|^2 | \al(t) \ra_1 {~}_1\la \be(t) | \otimes | \be(t) \ra_2 {~}_2\la \al(t) |
\nonumber\\
& \qquad & \qquad
\pm |f(t)|^2 | \be(t) \ra_1 {~}_1\la \al(t) | \otimes | \al(t) \ra_2 {~}_2\la \be(t) |
\bigg).
\label{eq: rhopmt}
\end{eqnarray}
Effect of statistics on the relative entropy of coherence is an interesting question. Here to simplify the calculations we compute this quantity for {\it reduced single-particle states} obtained by performing partial trace over the two-particle state (\ref{eq: rhopmt}). 
In this way we get
\begin{eqnarray} 
\rho_{\pm, \text{sp}}(t) &=& 
\mathcal{N}^2_{\pm} \bigg( | \al(t) \ra \la \al(t) | 
+ | \be(t) \ra \la \be(t) | \pm \la \al(t) | \be(t) \ra |f(t)|^2 | \al(t) \ra \la \be(t) | 
\nonumber \\ 
& \qquad & \qquad \pm \la \be(t) | \al(t) \ra |f(t)|^2 | \be(t) \ra \la \al(t) |
\bigg).
\label{rhopm-sp}
\end{eqnarray}
The corresponding classical one-particle state recasts
\begin{eqnarray} \label{rhoMB-sp}
\rho_{\MB, \text{sp}}(t) &=& 
\frac{1}{2} ( | \al(t) \ra \la \al(t) | + | \be(t) \ra \la \be(t) | ).
\end{eqnarray}
In figures \ref{fig: crrhosptime} we have numerically studied variation of the relative entropy of coherence of reduced states with time for a given value of damping constant but for two different values of $ \al $; taking one particle states as $ |\al \ra $ and $ |\be\ra = |- \al \ra $.
As one sees the coherence of fermionic states are less than those of two other statistics. Coherence of bosonic state is less than that of the classical state for $ \al = - \be = \frac{1}{ \sqrt{2} } $ while this reverses for the higher value $ \al = - \be = 1 $. Figure \ref{fig: crrhospalpha2} displays the coherence versus $ |\al|^2 $ for different statistics in the non-dissipative case. 
Note that one-particle states have been taken $|\al\ra$ and $|-\al \ra$. 
Separation between the representative points of these one-particle coherent states on the complex plane increases with $ | \al | $. Thus the overlap between the states decreases as $ | \al | $ increases. Negligible overlap of one-particle states implies negligible difference between different statistics. Thus, as the separation of one-particle states increases, results for all statistics approach as the figure illustrates. 

%

%
%

%
%

In the following subsections, we will study two interesting quantities. First, the MMS between particles for different statistics and second, the joint detection probability of both particles by an extended detector. The latter, which was previously considered for free particles for both the non-dissipative dynamics \cite{MaGr-EPJD-2014}, in the framework of the usual Schr\"{o}dinger equation, and for the dissipative one \cite{MoMi-EPJP-2020} in the context of both the Caldirola-Kanai time-dependent Hamiltonian and the master equation in the Caldeira-Leggett framework displays the unexpected boson anti-bunching and fermion bunching. Here, we extend this study to the case of non-free particles, i.e., DHOs obeying the dynamical equation \ref{eq: master}.

\subsection{ Mean square separation }

The mean square separation between particles is given by
\begin{equation}
\la ( \hat{x}_1 - \hat{x}_2 )^2 \ra_{\pm} = \tr[ ( \hat{x}_1 - \hat{x}_2 )^2 \hat{\rho}_{\pm}(t) ]
= \int_{-\infty}^{\infty} dx_1 \int_{-\infty}^{\infty} dx_2 ~ \la x_1, x_2 | ( \hat{x}_1 - \hat{x}_2 )^2 \hat{\rho}_{\pm}(t) | x_1, x_2 \ra 
\end{equation} 
which using (\ref{eq: rhopmt}) recasts 
\begin{eqnarray} 
\la ( \hat{x}_1 - \hat{x}_2 )^2 \ra_{\pm} &=&
2 \mathcal{N}_{\pm}^2 \bigg( \la ( \hat{x}_1 - \hat{x}_2 )^2 \ra_{\MB} 
\mp 2 |f(t)|^2 | \la \hat{x} \ra_{\al(t) \be(t)} |^2 
\nonumber \\
& \qquad & \qquad ~~~ \pm 
2 |f(t)|^2 \re \left\{ \la \hat{x}^2 \ra_{\al(t) \be(t)} \la \be(t) | \al(t) \ra \right\} 
\bigg) 
\label{eq: mss}
\end{eqnarray}
where
\begin{eqnarray} \label{eq: mssMB}
\la ( \hat{x}_1 - \hat{x}_2 )^2 \ra_{\MB} = \la \hat{x}^2 \ra_{\al(t)\al(t)} + \la \hat{x}^2 \ra_{\be(t)\be(t)} - 2 \la \hat{x} \ra_{\al(t)\al(t)} \la \hat{x} \ra_{\be(t) \be(t)} 
\end{eqnarray}
gives the MSS between distinguishable particles obeying the Maxwell-Boltzmann (MB) statistics, and 
\begin{eqnarray}
\la \cdots \ra_{\al(t) \be(t)} &=& \la \al(t) | \cdots | \be(t) \ra .
\end{eqnarray}

By taking $\be = - \al$ and $\al$ a real number, one obtains
\begin{eqnarray}
\la ( \hat{x}_1 - \hat{x}_2 )^2 \ra_{\MB} &=& 1 + 8 \al^2 e^{-\ga_0 t} \cos^2(t)  \label{eq: mss_MB}
\\
\la ( \hat{x}_1 - \hat{x}_2 )^2 \ra_{\pm} &=&
\frac{1}{1 \pm e^{- 4 \al^2} } 
\bigg( 1 + 8 \al^2 e^{-\ga_0 t} \cos^2(t) \pm e^{-4\al^2}( 1 - 8 \al^2 e^{-\ga_0 t} \sin^2(t) ) 
\bigg) 
\nonumber \\
\label{eq: mss_pm}
\end{eqnarray}
from which one gets
\begin{numcases}~
\la ( \hat{x}_1 - \hat{x}_2 )^2 \ra_{\FD} - \la ( \hat{x}_1 - \hat{x}_2 )^2 \ra_{\MB}
= 8 \al^2 e^{-\ga_0 t}  \frac{1}{-1+e^{ 4 \al^2}} , \\
\la ( \hat{x}_1 - \hat{x}_2 )^2 \ra_{\MB} - \la ( \hat{x}_1 - \hat{x}_2 )^2 \ra_{\BE}
= 8 \al^2 e^{-\ga_0 t}  \frac{1}{1+e^{ 4 \al^2}} , \\
\la ( \hat{x}_1 - \hat{x}_2 )^2 \ra_{\FD} - \la ( \hat{x}_1 - \hat{x}_2 )^2 \ra_{\BE}
= 8 \al^2 e^{-\ga_0 t} \frac{1}{\sinh(4\al^2)} .
\end{numcases}
These relations show that MSS for identical spinless fermions is greater than that of distinguishable particles which is itself greater than the MSS for identical spinless bosons. 
At long times MSS takes the same value $\hb / m \om_0 $ for all kind of particles revealing unimportance of quantum statistics in this limit. 
MSS in a given time does not have a regular behaviour with the relaxation rate for bosons. This can be seen from the $\ga_0-$derivative of Eqs. (\ref{eq: mss_MB}) and (\ref{eq: mss_pm})
\begin{numcases}~
\frac{d}{d\ga_0}\la ( \hat{x}_1 - \hat{x}_2 )^2 \ra_{\MB} = 
- 8 \al^2 t ~e^{-\ga_0 t} \cos^2(t)
\\
\frac{d}{d\ga_0} \la ( \hat{x}_1 - \hat{x}_2 )^2 \ra_{\pm} =
- 8 \al^2 t ~ e^{-\ga_0 t} 
\frac{ e^{ 4 \al^2} \cos^2(t) \mp \sin^2(t) }{ \pm 1 + e^{ 4 \al^2} }
\end{numcases}
showing that MSS always decreases with relaxation for both distinguishable particles and identical spinless fermions but this is not true for identical spinless bosons. This can be easily seen for instance for times $ t = (n-1/2)\pi $ with $n$ a natural number.

\subsection{Joint detection probability: boson anti-bunching and fermion bunching }

To evaluate the simultaneous detection probability we need diagonal elements of the density matrix,
\begin{eqnarray} 
\rho_{\pm}(x_1, x_2, t) &=& \la x_1, x_2 | \rho_{\pm}(t) | x_1, x_2 \ra 
\nonumber \\ 
&=& 2 \mathcal{N}^2_{\pm} \left[ \rho_{\MB}(x_1, x_2, t) \pm |f(t)|^2 
\re\{ \psi_{\al(t)}(x_1) \psi^*_{\be(t)}(x_1) \psi_{\be(t)}(x_2) \psi^*_{\al(t)}(x_2)\} \right]
\nonumber\\
\end{eqnarray}
where,
\begin{eqnarray} \label{eq: rhoMB}
\rho_{\MB}(x_1, x_2, t) &=& \frac{1}{2} \left[  
|\psi_{\al(t)}(x_1)|^2 |\psi_{\be(t)}(x_2)|^2 + |\psi_{\be(t)}(x_1)|^2 |\psi_{\al(t)}(x_2)|^2
\right]  
\end{eqnarray}
In comparison to the non-dissipative dynamics one sees that $ |f(t)|^2 $, multiplier of the interference term, is responsible for decoherence i.e., loss of indistinguishability \cite{MoMi-EPJP-2022}. 
Then the ratio of simultaneous detection probability of indistinguishable particles to the distinguishable ones, by an extended detector with width $2d$ located at the origin, is computed to give 
\begin{eqnarray} 
p_{\pm}(t) &=& 
\frac{ p_{\substack{\BE \\ \FD}}(t)
}{ p_{\MB}(t) }= 
\frac{ \int_{-d}^{d} dx_1 \int_{-d}^{d} dx_2 ~ \rho_{\pm}(x_1, x_1; x_2, x_2, t)  }{ \int_{-d}^{d} dx_1 \int_{-d}^{d} dx_2 ~ \rho_{\MB}(x_1, x_1; x_2, x_2, t) } \\
&=&
2 \mathcal{N}_{\pm}^2
\left\{  
1 \pm |f(t)|^2 \frac{ \left| \int_{-d}^{d} dx ~ \psi_{\al(t)}(x) \psi^*_{\be(t)}(x) \right|^2 }
{ \int_{-d}^{d} dx ~ |\psi_{\al(t)}(x)|^2 \int_{-d}^{d} dx ~ |\psi_{\be(t)}(x)|^2 }
\right\} \label{eq: detprob}
\end{eqnarray}

Taking both $ \al $ and $ \beta $ real; and $ \beta = - \al $ one obtains
\begin{eqnarray*}
\int_{-d}^{d} dx ~ |\psi_{\al(t)}(x)|^2 &=& \int_{-d}^{d} dx ~ |\psi_{\be(t)}(x)|^2
\\ 
&=& \frac{1}{2} \left\{ \erf \left[ d+\sqrt{2} \al e^{ -\ga_0  t /2 } \cos (t) \right] + \erf \left[ d-\sqrt{2} \al e^{ - \ga_0 t /2 } \cos (t) \right] \right\}
\end{eqnarray*}
and 
\begin{eqnarray*}
\int_{-d}^{d} dx ~ \psi_{\al(t)}(x) \psi^*_{\be(t)}(x) &=& 
\frac{ e^{-4 \al^2 e^{ - \ga_0 t }} }{4} \bigg\{  
\erf \left[ d + i \sqrt{2} \al e^{ - \ga_0 t /2 } \sin (t) \right]
\\
& \qquad & \qquad \qquad \qquad
+ \erf \left[ d - i \sqrt{2} \al e^{ - \ga_0 t /2 } \sin (t) \right]
\bigg\} \nonumber
\end{eqnarray*}
where $\erf[x]$ displays error function.


%
\begin{figure} 
\centering
\includegraphics[width=12cm,angle=-0]{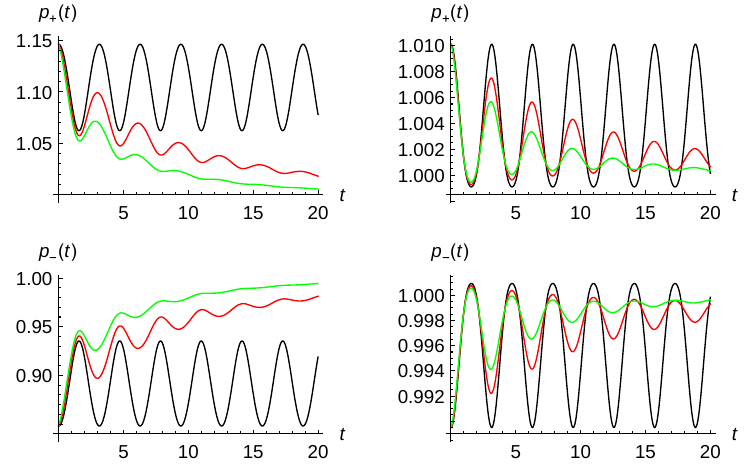}
\caption{(Color online)
Relative simultaneous detection probability $ p_+(t) = \frac{ p_{\BE}(t) }{ p_{\MB}(t) } $ (top plots) for two identical spinless bosons and $ p_-(t) = \frac{ p_{\FD}(t) }{ p_{\MB}(t) } $ (bottom plots) for two identical spinless fermions measured by a detector with a width $ 2d = 2 $ (left panels) and $ 2d = 4 $ (right panels) located at the origin for $ \al = -\be = 1 $ and different values of relaxation rate: $ \ga = 0 $ (black curves), $ \ga = 0.05 $ (red curves) and $ \ga = 0.1 $ (green curves).
}
\label{fig: detprob} 
\end{figure}
%

In figure \ref{fig: detprob}, relative joint detection probability of two identical particles, by a wide detector with a width $ 2d = 2 $ (left panels) and $ 2d = 4 $ (right panels) located at the origin, has been plotted for bosons (top plots) and fermions (bottom plots).
One sees that for $ 2d = 4 $ there are intervals of time for which $ p_+(t) < 1 $ while $ p_-(t) > 1 $. This is the hallmark of bosons anti-bunching and fermions bunching which has already been reported for zero external potential \cite{MaGr-EPJD-2014, MoMi-EPJP-2020}. Note that with dissipation, this effect disappears as time elapses.

It is notable that the impact of quantum entanglement on chaotic and ordered Bohmian trajectories has been investigated by using a two-qubit system composed of two {\it non-dissipative} coherent states \cite{TzCoEf-PS-2019}; and real and complex quantum trajectories resulting from {\it Klauder coherent states} have been computed in \cite{DF-PRA-2013}.


\section{Concluding remarks} \label{sec: conclude}

Although decoherence in systems of DHOs is an old problem \cite{DF-PRD-1989, DF-PRD-1994}, it is still an active field of research \cite{MoMi-JPC-2018, Po-JPA-2020}. Here, we addressed the problem through a novel approach by considering loss of quantum coherence with dissipation.
Different aspects of DHOs were considered. Evolution of relative entropy coherence of a coherent state was considered for different values of relaxation rate. Coherence decreases with time, and the rate of decrement is higher for higher values of damping constant. Furthermore, coherence of a superposition of two coherent states, a cat state, and also a superposition of two cat states were studied as a function of separation between two superposed states in the complex plane.
Density plots of the probability density in the position space were plotted for a cat state, and also the corresponding Bohmian trajectories were computed. 
Then, a system of two identical DHOs was considered and the relative entropy coherence for the reduced single-particle states was computed. It is seen that coherence for fermionic states is less than that of bosonic ones. 
This can be utilized as an indicator to distinguish between different statistics.
Furthermore, this result shows that when DHOs are used, the symmetric states are more useful as a resource for certain quantum information and computation protocols than the antisymmetric ones. 
%
Two interesting quantities, the MSS between particles and the joint detection probability of both particles by a non-ideal wide detector, were studied for different statistics. MSS for fermions is always greater than that of distinguishable particles, which itself is higher than the MSS for bosons. Furthermore, (anti)-bunching of (bosons) fermions is seen in this interacting system. 
As far as we found, this unexpected effect has already been reported for {\it only} free identical particles in the context of closed isolated systems obeying the Schr\"odinger equation \cite{MaGr-EPJD-2014}; and also in the context of open systems in the formalism of both Caldirola-Kanai and Caldeira-Leggett \cite{MoMi-EPJP-2020} but here we extended the study to systems of identical DHOs.


\section*{Acknowledgements}
Support from the University of Qom is acknowledged. 

%

%
%


%
\end{document}